\documentclass[a4paper]{article}

\usepackage[margin=1in]{geometry}

\usepackage[T1]{fontenc}
\newcommand{\changefont}[3]{
\fontfamily{#1} \fontseries{#2} \fontshape{#3} \selectfont}

\changefont{ptm}{m}{n}

\usepackage{setspace} 

\usepackage{graphicx}    % needed for including graphics e.g. EPS, PS

\usepackage{amsfonts}
\usepackage{amssymb}
\usepackage{graphics}
\usepackage{mathrsfs}
\usepackage{color}
\newtheorem{remark}{Remark}[section]

\newtheorem{theorem}{Theorem}[section]

\newtheorem{definition}{Definition}[section]

\long\def\symbolfootnote[#1]#2{\begingroup%
\def\thefootnote{\fnsymbol{footnote}}\footnote[#1]{#2}\endgroup} 

\begin{document}

%\begin{frontmatter}
%

\begin{center}
\Large \textbf{Unpredictable solutions of differential equations}
\end{center}

\begin{center}
\normalsize \textbf{Marat Akhmet$^{a,}\symbolfootnote[1]{Corresponding Author Tel.: +90 312 210 5355,  Fax: +90 312 210 2972, E-mail: marat@metu.edu.tr}$, Mehmet Onur Fen$^b$} \\
\vspace{0.2cm}
\textit{\textbf{\footnotesize$^a$Department of Mathematics, Middle East Technical University, 06800 Ankara, Turkey}} \\
\textit{\textbf{\footnotesize$^b$Basic Sciences Unit, TED University, 06420 Ankara, Turkey}}
\vspace{0.1cm}
\end{center}

\vspace{0.3cm}

\begin{center}
\textbf{Abstract}
\end{center}

\noindent\ignorespaces
We apply the topology of convergence on compact sets to define unpredictable functions \cite{Akhmet17,Akh18}. The topology is metrizable and easy for applications with integral operators. To demonstrate the effectiveness of the approach, the existence and uniqueness of the unpredictable solution for a delay differential equation is proved. As a corollary of the theorem, a similar assertion for a quasilinear ordinary differential equation is formulated. Examples with simulations illustrate the obtained results. 

\vspace{0.2cm}
 
\noindent\ignorespaces \textbf{Keywords:} Unpredictable solutions; Unpredictable sequences; Topology of convergence on compact subsets; Poincar\'e chaos; Delay differential equations; Ordinary differential equations

\vspace{0.6cm}

\section{Introduction and preliminaries}     

From the applications point of view, the theory of differential equations focuses on equilibria, periodic, and almost periodic oscillations. They meet the needs of any real world problem related to mechanics, electronics, economics, biology, etc., if one searches for regular and stable dynamics of an isolated motion. However, they are not sufficient for many modern and perspective demands of robotics, computer techniques, and the internet, and chaotic dynamics comprise constructive properties for applications. This is the reason why it is important to join the power of deterministic  chaos with the immensely rich source of methods for differential equations. We contributed to this in our papers \cite{Akh8,Akh12} and the book \cite{Akh14}, where a method of replication of chaos has been developed. It consists of the verification of ingredients of chaos such as sensitivity, transitivity, proximality and existence of infinitely many unstable regular motions \cite{Devaney87,Li75,Lorenz61,Wiggins88} for solutions of an equation with chaotic perturbation. This approach gives a very effective instrument for application of the accumulated knowledge in chaos research. Nevertheless, we are not glad with the necessity to check the presence of several ingredients. Therefore, in our opinion, unpredictable functions have become a panacea for simplification of chaos analysis through differential equations. 

In this paper, another step in adaptation of unpredictable functions to the theory of differential equations has been made. We apply convergence on compact subsets of the real axis to determine unpredictable functions for two reasons. The first reason is that the topology is easily metrizable, in particular, to the metric for Bebutov dynamical system \cite{Sell71}, and consequently, the unpredictable functions and solutions immediately imply the Poincar\'e chaos presence according  to  our results in \cite{Akhmet16}. The second one is the easy verification of the convergence. Thus, the present study is useful for the theory of differential equations as well as chaos researches. For the construction of unpredictable functions we have applied the significant results on the equivalence of discrete dynamics obtained in papers \cite{Shi04,Shi07}.

Let us introduce the following definition. 

\begin{definition} \label{definition_unpredictable}
A uniformly continuous and bounded function $\varphi: \mathbb R  \to \mathbb R^m$ is unpredictable if there exist positive numbers $\epsilon_0,$ $\delta,$ and sequences $\left\{t_n\right\},$ $\left\{u_n\right\}$ both of which diverge to infinity such that $\left\|\varphi(t+t_n)-\varphi(t)\right\| \to 0$ as $n \to \infty$ on compact subsets of $\mathbb R,$ and $\left\|\varphi(t+t_n)-\varphi(t)\right\| \ge \epsilon_0$ for each $t\in [u_n-\delta, u_n+\delta]$ and $n\in\mathbb N.$
\end{definition}

To create Poincar\'e chaos \cite{Akhmet16}, uniform continuity is not a necessary condition for an unpredictable function $\varphi(t)$, and instead of the condition $\left\|\varphi(t+t_n)-\varphi(t)\right\| \ge \epsilon_0$ for each $t\in [u_n-\delta, u_n+\delta]$ and $n\in\mathbb N,$ one can request that $\left\|\varphi(t_n+u_n)-\varphi(u_n)\right\| \geq \epsilon_{0}$ for each $n \in \mathbb N.$ For the needs of verification of theorems on the existence of unpredictable solutions of differential equations we apply Definition \ref{definition_unpredictable},  but for the future studies the  following  definition may also be beneficial.

\begin{definition} \label{definition_unpredictable1}
A continuous and bounded function $\varphi: \mathbb R \to \mathbb R^m$ is unpredictable if there exist positive numbers $\epsilon_0,$ $\delta,$ and sequences $\left\{t_n\right\},$ $\left\{u_n\right\}$ both of which diverge to infinity such that $\left\|\varphi(t+t_n)-\varphi(t)\right\| \to 0$ as $n \to \infty$ on compact subsets of $\mathbb R,$ and $\left\|\varphi(t_n+u_n)-\varphi(u_n)\right\| \ge \epsilon_0$ for each $n\in\mathbb N.$
\end{definition}

The main object of the present study is the following system of delay differential equations,
\begin{eqnarray} \label{main_delay_system}
x'(t)=Ax(t)+f(x(t-\tau)) + g(t),
\end{eqnarray}
where $\tau$ is a positive number, the eigenvalues of the matrix $A \in \mathbb R^{m \times m}$ have negative real parts, $f:\mathbb R^m \to \mathbb R^m$ is a continuous function, and $g:\mathbb R \to \mathbb R^m$ is a uniformly continuous and bounded function. Our purpose is to prove that system (\ref{main_delay_system}) possesses a unique unpredictable solution which is uniformly exponentially stable, provided that the function $g(t)$ is unpredictable in accordance with Definition \ref{definition_unpredictable}.

\section{Unpredictable solutions}

In the remaining parts of the paper, we will make use of the usual Euclidean norm for vectors and the norm induced by the Euclidean norm for square matrices \cite{Horn92}.  

Since the eigenvalues of the matrix $A$ in system (\ref{main_delay_system}) have negative real parts, there exist numbers $K \ge 1$ and $\omega >0$ such that $\left\|e^{At}\right\| \le K e^{-\omega t}$ for $t\ge 0.$

The following conditions are required.

\begin{enumerate}
\item[\textbf{(C1)}] There exists a positive number $M_{f}$ such that $\displaystyle \sup_{x\in\mathbb R^{m}} \left\|f(x)\right\| \le M_f;$ 
\item[\textbf{(C2)}] There exists a positive number $L_{f}$ such that $\left\|f(x_{1})-f(x_{2})\right\| \le L_{f} \left\|x_{1}-x_{2}\right\|$ for all $x_{1},x_{2} \in \mathbb R^{m};$
\item[\textbf{(C3)}] $\omega - 2 K L_{f} e^{\omega \tau /2} > 0.$
\end{enumerate}

The next theorem is concerned with the existence of an unpredictable solution of system (\ref{main_delay_system}).

\begin{theorem} \label{main_thm1}
Suppose that conditions $(C1)-(C3)$ are valid. If the function $g(t)$ is unpredictable, then system (\ref{main_delay_system}) possesses a unique uniformly exponentially stable unpredictable solution.
\end{theorem}

\noindent \textbf{Proof.} Under the conditions $(C1)-(C3)$, one can verify using the techniques for delay differential equations \cite{Driver77} that there exists a unique solution $\phi(t)$ of (\ref{main_delay_system}) which is bounded on the whole real axis and satisfies the relation
\begin{eqnarray*} \label{integral_eqn}
\phi(t) = \displaystyle \int_{-\infty}^{t} e^{A(t-s)} [f(\phi(s-\tau)) + g(s)] ds.
\end{eqnarray*}
It is clear that $\displaystyle \sup_{t\in\mathbb R} \left\|\phi(t)\right\| \le M_{\phi},$ where $M_{\phi}=\displaystyle \frac{K(M_f+M_g)}{\omega}$ and $M_g=\displaystyle \sup_{t\in\mathbb R} \left\|g(t)\right\|.$ According to results of \cite{Driver77}, the solution $\phi(t)$ is uniformly exponentially stable. We will show that the solution $\phi(t)$ is unpredictable.

Since $g(t)$ is an unpredictable function, there exist a positive number $\epsilon_0$ and sequences $\left\{t_n\right\},$ $\left\{u_n\right\}$ both of which diverge to infinity such that $\left\|g(t+t_n)-g(t)\right\| \to 0$ as $n \to \infty$ on compact subsets of $\mathbb R,$ and $\left\|g(t_n+u_n)-g(u_n)\right\| \ge \epsilon_0$ for each $n\in\mathbb N.$

Fix an arbitrary $\epsilon>0,$ and denote $R_1= \displaystyle \frac{2 \omega M_{\phi} K}{\omega - 2 K L_f e^{\omega \tau /2}},$ $R_2=\displaystyle \frac{K}{\omega-K L_f}.$ Condition $(C3)$ implies that both $R_1$ and $R_2$ are positive numbers. Take a positive number $\gamma$ satisfying $\gamma < \displaystyle \frac{1}{R_1+R_2},$ and suppose that $E$ is a positive number such that $E \ge \displaystyle \frac{2}{\omega} \ln \left(\frac{1}{\gamma \epsilon}\right).$

Let $[\alpha, \beta]$ be a compact subset of $\mathbb R$, where $\beta > \alpha.$ There exists a natural number $n_{0}$ such that for each $n \ge n_{0}$ the inequality $\left\|g(t+t_n) - g(t)\right\| < \gamma \epsilon$ holds for $t \in [\alpha -E, \beta].$ 

Fix an arbitrary natural number $n \ge n_0,$ and define the function $z(t)=\phi(t)-\phi(t+t_n).$ For $t\ge \alpha-E,$ $z(t)$ satisfies the relation
\begin{eqnarray*}
 z(t)& = & e^{A(t-\alpha+E)} [ \phi(\alpha-E) - \phi(t_n+\alpha-E) ] \\
&& + \displaystyle \int_{\alpha-E}^{t} e^{A(t-s)} [f(z(s-\tau) + \phi(s+t_n-\tau)) - f(\phi(s+t_n-\tau))] ds \\
&& + \displaystyle \int_{\alpha-E}^{t} e^{A(t-s)} [g(s)-g(s+t_n)] ds.
\end{eqnarray*}
 
Let us denote by $\mathscr{C}$ the set of continuous functions $z(t)$ defined on $\mathbb R$ such that $$\left\|z(t)\right\| \le R_{1} e^{-\omega (t-\alpha +E)/2 } +R_2 \gamma \epsilon$$ for $\alpha-E-\tau \le t \le \beta$ and $\left\|z\right\|_{\infty} \le 2K\displaystyle \left(M_{\phi} + \frac{M_{f}+M_{g}}{\omega}\right),$ where $\left\|z\right\|_{\infty}=\displaystyle \sup_{t\in\mathbb R} \left\|z(t)\right\|$.

Define on $\mathscr{C}$ the operator $\Pi$ as
\begin{eqnarray*}
\Pi z(t)= \left\{\begin{array}{ll}   \phi(t) - \phi(t+t_n),  ~ t < \alpha-E, \\
 e^{A (t-\alpha+E)} \left[ \phi(\alpha-E) - \phi(t_n+\alpha-E)  \right]  + \displaystyle \int_{\alpha-E}^{t} e^{A(t-s)} [g(s)  - g(s+t_n) ] ds \\
 +\displaystyle \int_{\alpha-E}^t e^{A(t-s)} [f(z(s-\tau) + \phi(s+t_n-\tau)) - f(\phi(s+t_n-\tau))]  ds, ~ t\ge \alpha-E. 
\end{array}\right.
\end{eqnarray*}

First of all, we will show that $\Pi(\mathscr{C}) \subseteq \mathscr{C}.$ If $z(t)$ belongs to $\mathscr{C},$ then we have for $t\in [\alpha-E,\beta]$ that 
\begin{eqnarray*}
 \left\|\Pi z(t)\right\| &\le & K e^{-\omega(t-\alpha+E)} \left\|\phi(\alpha-E) - \phi(t_n+\alpha-E)\right\| + \displaystyle \int_{\alpha-E}^{t} K\gamma \epsilon e^{-\omega(t-s)} ds \\
&& +\displaystyle \int_{\alpha-E}^{t} K L_{f} e^{-\omega (t-s)} \left\|z(s-\tau)\right\| ds \\
&<&  \left( 2 M_{\phi} K + \frac{2KL_{f}R_{1} e^{\omega \tau /2}}{\omega} \right) e^{-\omega (t-\alpha+E)/2}  + \displaystyle \frac{K\gamma \epsilon (1+L_{f}R_{2})}{\omega} \\
&=& R_{1} e^{-\omega(t-\alpha+E)/2} + R_{2} \gamma \epsilon.
\end{eqnarray*}
The inequality $\left\|\Pi z(t)\right\| < R_{1} e^{-\omega(t-\alpha+E)/2} + R_{2} \gamma \epsilon$ is valid also for $t\in[\alpha-E-\tau,\alpha-E)$ since $R_{1}>2M_{\phi}.$ On the other hand, if $z(t)$ belongs to $\mathscr{C},$ then one can confirm that $\left\|\Pi z\right\|_{\infty} \le 2K\left(M_{\phi} + \displaystyle \frac{M_{f}+M_{g}}{\omega}\right).$ Hence, $\Pi(\mathscr{C}) \subseteq \mathscr{C}.$

Now,  let us take two functions $z(t),\overline{z}(t) \in \mathscr{C}.$ Clearly, $\Pi z(t)-\Pi \overline{z}(t)=0$ for each $t<\alpha-E.$ It can be verified for $t \geq \alpha-E$ that
\begin{eqnarray*}
\left\|\Pi z(t) - \Pi \overline{z}(t)\right\| & \leq & \displaystyle \int_{\alpha-E}^{t} K L_{f} e^{-\omega (t-s)} \left\|z(s-\tau) - \overline{z}(s-\tau)\right\| ds \\
&\le & \displaystyle \frac{K L_{f}}{\omega} \left(1-e^{-\omega (t-\alpha+E)}\right) \sup_{t \geq \alpha-E-\tau} \left\|z(t) - \overline{z}(t)\right\|. 
\end{eqnarray*}
The last inequality yields $\left\|\Pi z - \Pi \overline{z}\right\|_{\infty} \leq \displaystyle \frac{K L_{f}}{\omega} \left\|z-\overline{z}\right\|_{\infty}.$ Thus, the operator $\Pi$ is contractive by means of condition $(C3).$

According to the uniqueness of solutions, $z(t)=\phi(t)-\phi(t+t_{n})$ is the unique fixed point of the operator $\Pi.$ Therefore,  $\left\|\phi(t+t_{n})-\phi(t)\right\| \leq R_{1} e^{\omega (t-\alpha+ E) / 2} + R_{2} \gamma \epsilon$ for $t \in [\alpha -E, \beta].$

Since the number $E$ is sufficiently large such that $E \ge \displaystyle \frac{2}{\omega} \ln \left(\frac{1}{\gamma \epsilon}\right),$ we have for $t\in[\alpha,\beta]$ that
\begin{eqnarray*}
\left\|\phi(t+t_{n})- \phi(t)\right\| \leq (R_{1} + R_{2}) \gamma \epsilon < \epsilon.
\end{eqnarray*}
Thus, $\left\|\phi(t+t_{n})-\phi(t)\right\| \to 0$ as $n \to \infty$ for $t \in[\alpha,\beta].$

In the remaining part of the proof, we will show the existence of a sequence $\left\{\eta_n\right\},$ $\eta_n \to \infty$ as $n\to \infty,$ and positive numbers $\overline{\epsilon}_{0},$ $\delta$ such that $\left\|\phi(t+t_n) - \phi(t)\right\| \ge \overline{\epsilon}_{0}$ for $t \in \big[\eta_{n} -\delta, \eta_{n}+\delta \big].$

Fix a natural number $n.$ Since the function $g(t)$ is uniformly continuous, there exists a positive number $\widetilde{\delta},$ which is independent of $t_{n}$ and $u_{n},$ such that both of the inequalities
$$
\left\|g(t+t_{n}) - g(t_{n}+u_{n}) \right\| \leq \frac{\epsilon_{0}}{4m}
$$ 
and
$$
\left\|g(t)-g(u_{n})\right\| \leq \frac{\epsilon_{0}}{4m}
$$
hold for $t \in [u_{n}-\widetilde{\delta},u_{n}+\widetilde{\delta}].$

Suppose that $g(t)=(g_{1}(t), g_{2}(t), \ldots, g_{m}(t)),$ where each $g_i(t),$ $1\leq i \leq m,$ is a real valued function.  One can confirm that there exists an integer $j_{n},$ $1 \leq j_{n} \leq m,$ such that 
$$
\left|g_{j_{n}}(t_{n}+u_{n}) - g_{j_{n}}(u_{n}) \right| \geq \displaystyle \frac{\epsilon_{0}}{m}.
$$
Therefore, using the last inequality, we obtain for $t \in [u_{n} - \widetilde{\delta}, u_{n} + \widetilde{\delta}]$ that 
\begin{eqnarray} \label{proof_ineqq1}
\left|g_{j_{n}}(t+t_{n})-g_{j_{n}}(t)\right| &\geq& \left|g_{j_{n}}(t_{n}+u_{n})-g_{j_{n}}(u_{n})\right| - \left|g_{j_{n}}(t+t_{n})-g_{j_{n}}(t_{n}+u_{n})\right| \nonumber\\ 
&& -  \left|g_{j_{n}}(t)-g_{j_{n}}(u_{n})\right| \nonumber\\ 
&\geq & \frac{\epsilon_{0}}{2m}.
\end{eqnarray}

There exist numbers $s_1^n, s_2^n, \ldots, s_m^n \in \Big[u_n-\widetilde{\delta},u_n+\widetilde{\delta}\Big]$ such that
$$
\left\|\displaystyle \int_{u_n-\widetilde{\delta}}^{u_n+\widetilde{\delta}} \left(g(s+t_{n})-g(s)\right) ds \right\| = 2 \widetilde{\delta} \left[\sum_{i=1}^{m}\left(g_{i}(s_{i}^{n}+t_{n})-g_{i}(s_{i}^{n})\right)^{2}\right]^{1/2}.
$$
Accordingly, the inequality (\ref{proof_ineqq1}) implies that
$$
\left\|\displaystyle \int_{u_n-\widetilde{\delta}}^{u_n+\widetilde{\delta}} \left(g(s+t_{n})-g(s)\right) ds \right\| \geq 2 \widetilde{\delta} \left| g_{j_{n}}(s_{j_{n}}^{n}+t_{n})-g_{j_{n}}(s_{j_{n}}^{n}) \right| \geq \frac{\widetilde{\delta} \epsilon_{0}}{m}.
$$
Now, using the relation
\begin{eqnarray*}
\phi \left(t_{n}+u_{n}+\widetilde{\delta} \right)-\phi \left(u_{n}+\widetilde{\delta} \right) & = & \phi \left(t_{n}+u_{n}-\widetilde{\delta} \right)-\phi \left(u_{n}-\widetilde{\delta} \right) 
 + \displaystyle \int_{u_{n}-\widetilde{\delta}}^{u_{n}+\widetilde{\delta}} A [\phi(s+t_{n})-\phi(s)] ds  \\
 && + \displaystyle \int_{u_{n}-\widetilde{\delta}}^{u_{n}+\widetilde{\delta}} [ f(\phi(s+t_{n}-\tau)) - f(\phi(s-\tau)) ] ds \\
 && + \displaystyle \int_{u_{n}-\widetilde{\delta}}^{u_{n}+\widetilde{\delta}} [g(s+t_{n})-g(s)] ds,
\end{eqnarray*}
one can verify that
\begin{eqnarray*}
\left\|\phi \left(t_{n}+u_{n}+\widetilde{\delta} \right)-\phi \left(u_{n}+\widetilde{\delta} \right) \right\| & \geq & \displaystyle \frac{\widetilde{\delta} \epsilon_{0}}{m} -  \left(1+2\widetilde{\delta} \left\|A\right\| \right) \sup_{t \in [u_{n}-\widetilde{\delta}, u_{n}+\widetilde{\delta}]} \left\|\phi(t+t_{n})-\phi(t)\right\| \\
&& - 2 \widetilde{\delta} L_{f}  \sup_{t \in [u_{n}-\widetilde{\delta}-\tau, u_{n}+\widetilde{\delta}-\tau]} \left\|\phi(t+t_{n})-\phi(t)\right\|.
\end{eqnarray*}
Hence, we have 
$$
\sup_{t \in [u_{n}-\widetilde{\delta}-\tau, u_{n}+\widetilde{\delta}]} \left\|\phi(t+t_{n})-\phi(t)\right\| \geq \displaystyle \frac{\widetilde{\delta} \epsilon_{0}}{2m \left(1+\widetilde{\delta}\left\|A\right\| + \widetilde{\delta} L_{f} \right)}.
$$

Let $\eta_{n}$ be a point that belongs to the interval $\Big[u_{n}-\widetilde{\delta}-\tau, u_{n}+\widetilde{\delta} \Big]$ satisfying
$$
\sup_{t \in [u_{n}-\widetilde{\delta}-\tau, u_{n}+\widetilde{\delta}]} \left\|\phi(t+t_{n})-\phi(t)\right\| = \left\|\phi(t_{n}+\eta_{n})-\phi(\eta_{n})\right\|.
$$
Define the numbers 
$$
\overline{\epsilon}_{0}=\frac{\widetilde{\delta} \epsilon_{0}}{4 m \left(1+\widetilde{\delta} \left\|A\right\| + \widetilde{\delta} L_{f}\right)}
$$
and
$$
\delta= \frac{\widetilde{\delta} \epsilon_{0}}{8 m \left(1+\widetilde{\delta} \left\|A\right\| + \widetilde{\delta} L_{f}\right) \left(M_{\phi} \left\|A\right\| + M_{\phi} L_{f} + M_{g}\right)}.
$$
If $t$ belongs to the interval $[\eta_{n}-\delta, \eta_{n}+\delta],$ then it can be obtained that 
\begin{eqnarray*}
\left\|\phi(t+t_{n}) - \phi(t)\right\| & \geq & \left\|\phi(t_{n}+\eta_{n}) - \phi(\eta_{n})\right\| - \left|\displaystyle \int_{\eta_{n}}^{t} \left\|A\right\| \left\|\phi(s+t_{n}) - \phi(s)\right\| ds \right| \\
&& - \left|\displaystyle \int_{\eta_{n}}^{t} L_{f} \left\|\phi(s+t_{n}-\tau) - \phi(s-\tau)\right\| ds \right| \\
&& - \left|\displaystyle \int_{\eta_{n}}^{t} \left\|g(s+t_{n}) - g(s)\right\| ds \right| \\
& \geq & \displaystyle \frac{\widetilde{\delta} \epsilon_{0}}{2m\left(1+\widetilde{\delta} \left\|A\right\|+\widetilde{\delta}L_{f}\right)} - 2\delta \left(M_{\phi} \left\|A\right\| + M_{\phi} L_{f} + M_{g}\right) \\
& = & \overline{\epsilon}_{0}.
\end{eqnarray*}

Hence, $\left\|\phi(t+t_{n})-\phi(t)\right\| \geq \overline{\epsilon}_{0}$ for each $t$ from the intervals $[\eta_{n}-\delta, \eta_{n}+\delta],$ $n\in\mathbb N.$ Clearly, $\eta_{n} \to \infty$ as $n\to\infty.$  Consequently, the bounded solution $\phi(t)$ is unpredictable. $\square$

\begin{remark}
The result of Theorem \ref{main_thm1} is valid also for the case $\tau=0.$ More precisely, if $(C1),$ $(C2)$ are valid and $\omega - K L_{f} > 0,$ then the system
\begin{eqnarray*}  
x'(t)=Ax(t)+f(x(t)) + g(t),
\end{eqnarray*}
possesses a unique uniformly exponentially stable unpredictable solution provided that $g(t)$ is an unpredictable function.
\end{remark}

\section{Unpredictable sequences}

The definition of an unpredictable sequence is as follows.

\begin{definition} \label{unp_seq}
A bounded sequence $\left\{\kappa_i\right\}$ in a metric space $(X,d)$ is called unpredictable if there exist a positive number $\epsilon_{0}$ and  sequences $\left\{j_n\right\},$ $\left\{k_n\right\}$ of positive integers both of which diverge to infinity such that $d(\kappa_{i+j_n}, \kappa_i) \to 0$ as $n \to \infty$ for each $i$ in bounded intervals of integers and $d(\kappa_{j_n+k_n}, \kappa_{k_n}) \geq \epsilon_{0}$ for each $n\in\mathbb N.$
\end{definition}

Let us consider the space $\displaystyle \Sigma_2=\left\{s=(s_0s_1s_2\ldots) \ | \ s_j =0 \ \textrm{or} \ 1 \right\}$ of infinite sequences of $0$'s and $1$'s with the metric $$d(s,t)=\displaystyle \sum_{k=0}^{\infty} \frac{\left|s_k-t_k\right|}{2^k},$$ where $s=(s_0s_1s_2\ldots),$ $t=(t_0t_1t_2\ldots)\in \Sigma_2.$ The Bernoulli shift $\sigma: \Sigma_2 \to \Sigma_2$ is defined as $\sigma(s_0s_1s_2\ldots)=(s_1s_2s_3\ldots).$ The map $\sigma$ is continuous and $\Sigma_2$ is a compact metric space \cite{Devaney87,Wiggins88}.

Through the proof of Lemma 3.1. \cite{Akh18}, we constructed an element $s^{**}=(s_0^{**}s_1^{**}s_2^{**}\ldots)$ of $\Sigma_2$ which is unpredictable in the sense of Definition \ref{unp_seq} by placing all blocks of $0$'s and $1$'s in a specific order without any repetitions and extending it to the left hand side by appropriately choosing the terms $s^{**}_i$ for negative values of $i.$

Now, we take into account the logistic map
\begin{eqnarray} \label{logistic_map_example}
\zeta_{i+1}= F_{\mu}(\zeta_{i}),
\end{eqnarray}
where $i \in \mathbb Z$ and $F_{\mu}(s)=\mu s (1-s).$ The interval $[0,1]$ is invariant under the iterations of (\ref{logistic_map_example}) for $\mu \in (0,4]$ \cite{Hale91}.

It was proved by Shi and Yu \cite{Shi07} that for each $\mu \in [3+(2/3)^{1/2}, 4],$ there exist a natural number $h_0 > 4$ and a Cantor set $\Lambda \subset [0,1]$ such that the map $F_{\mu}^{h_0}$ on $\Lambda$ is topologically conjugate to the Bernoulli shift $\sigma$ on $\Sigma_2.$ This gives immediately the proof of the following theorem.

\begin{theorem}
For each $\mu \in [3+(2/3)^{1/2},4],$ the logistic map (\ref{logistic_map_example}) possesses an unpredictable solution.
\end{theorem}

The next section is devoted to examples.

\section{Examples}

\subsection{Example 1}

Let $\{\zeta^{*}_{i}\},$ $i\in \mathbb Z,$ be an unpredictable solution of the logistic map (\ref{logistic_map_example}) with $\mu=3.91$ inside the unit interval $[0,1],$ and consider the differential equation
\begin{eqnarray}
v'(t)=-2v(t)+p(t), \label{system1_example}
\end{eqnarray}
where the function $p(t)$ is defined as $p(t)=\zeta^{*}_{i}$ for $t\in [i,i+1),$ $i\in\mathbb Z.$ Equation (\ref{system1_example}) is a hybrid system since its dynamics is governed by both a differential equation and a discrete one.

It can be confirmed that the function 
\begin{eqnarray}  \label{func_unprdctble_psi}
\psi(t)=\displaystyle \int_{-\infty}^{t} e^{-2(t-s)} p(s) ds
\end{eqnarray}
is the unique global exponentially stable solution of (\ref{system1_example}) which is bounded on the whole real axis such that $\displaystyle \sup_{t\in\mathbb R} \left|\psi(t)\right| \leq \frac{1}{2}.$ Moreover, $\psi(t)$ is uniformly continuous, since its derivative is bounded.

Because the sequence $\{\zeta^{*}_{i}\},$ $i\in \mathbb Z,$ is unpredictable, there exist a positive number $\epsilon_{0}$ and sequences $\{j_n\},$ $\{k_{n}\}$ both of which diverge to infinity such that $\left|\zeta^{*}_{i+j_{n}} - \zeta^{*}_{i}\right| \to 0$ as $n \to \infty$ for each $i$ in bounded intervals of integers and $\left|\zeta^{*}_{j_{n}+k_{n}} - \zeta^{*}_{k_{n}}\right| \geq \epsilon_{0}$ for each $n.$

Fix an arbitrary positive number $\epsilon,$ take an arbitrary compact subset $[\alpha, \beta] \subset \mathbb{R}.$ Suppose that $N$ is a sufficiently large positive integer satisfying $\displaystyle N \geq \frac{1}{2} \ln \left(\frac{3}{2 \epsilon}\right).$
There exists a natural number $n_{0}$ such that for each $n \geq n_{0}$ the inequality 
$$
\left|\zeta^{*}_{i+j_{n}}-\zeta^{*}_{i}\right| < \displaystyle \frac{2 \epsilon}{3}
$$ 
is valid for $i=\lfloor \alpha \rfloor -N, \lfloor \alpha \rfloor -N +1, \ldots, \lfloor \beta \rfloor,$ where $\lfloor \alpha \rfloor$ and $\lfloor \beta \rfloor$ denote the largest integers which are not greater than $\alpha$ and $\beta,$ respectively.

Fix a natural number $n \geq n_{0}.$ Using the relation
\begin{eqnarray*}
\psi(t+j_{n}) - \psi(t) &=& e^{-2(t-\lfloor \alpha \rfloor+N)} (\psi(\lfloor \alpha \rfloor-N+j_{n})-\psi(\lfloor \alpha \rfloor-N)) \\
& &+  \displaystyle \int_{\lfloor \alpha \rfloor-N}^{t} e^{-2(t-s)} [p(s+j_{n})-p(s)] ds,
\end{eqnarray*}
one can verify for $t\in [\lfloor \alpha \rfloor, \lfloor \beta \rfloor+1)$ that $\left|\psi(t+j_{n}) - \psi(t)\right|<\epsilon.$ Hence, $\psi(t+ j_{n}) \to \psi(t)$ as $n\to\infty$ on $[\alpha,\beta].$

On the other hand, one can show that $\displaystyle \sup_{t\in [k_{n},k_{n}+1]} \left|\psi(t+j_{n})-\psi(t)\right|\geq \frac{\epsilon_{0}}{4}$ for each $n\in\mathbb{N}.$ The last inequality implies that the function $\psi(t)$ is unpredictable.

\begin{figure}[ht] 
\centering
\includegraphics[width=14.0cm]{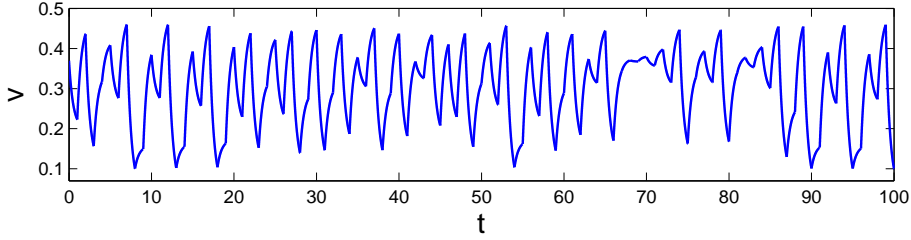}
\caption{Chaotic behavior of equation (\ref{system1_example}). The figure represents the solution of (\ref{system1_example}) with $v(0)=0.37$ and $\zeta_{0}=0.4.$}
\label{fig1}
\end{figure}

We depict in Figure \ref{fig1} the solution of (\ref{system1_example}) with initial data $v(0)=0.37$ and $\zeta_{0}=0.4.$ The choice of the parameter $\mu=3.91$ of the logistic map (\ref{logistic_map_example}) and the value of $\zeta_{0}=0.4$ are considered for shadowing in the paper \cite{Hammel87}. It is seen in  Figure \ref{fig1} that the dynamics of (\ref{system1_example}) is chaotic, and this supports that the bounded solution $\psi(t)$ is unpredictable.    

\subsection{Example 2}
In this example, we take into account the retarded differential equation
\begin{eqnarray}
x''(t)+4x'(t)+1.5x(t)+0.02x^2(t-0.1)=\psi(t),  \label{system2_example}
\end{eqnarray}
where $\psi(t)$ is the unpredictable solution of (\ref{system1_example}) defined by equation (\ref{func_unprdctble_psi}).

Using the variables $x_1(t)=x(t)$ and $x_2(t)=x'(t),$ equation (\ref{system2_example}) can be written as 
\begin{eqnarray} \label{example_delay_system}
&& x'_1(t)=x_2(t), \nonumber \\ 
&& x'_2(t)=-1.5x_1(t)-4x_2(t)-0.02 x_1^2(t-0.1)+\psi(t).
\end{eqnarray}

System (\ref{example_delay_system}) is in the form of (\ref{main_delay_system}) with $\tau=0.1,$ 
$f(x_{1},x_{2})=\left(0, -0.02 x_1^2\right),$ and
$A=\left( \begin{array}{cc}
0 & 1 \\
-1.5 & -4 \end{array} \right).$
The eigenvalues of the matrix $A$ are $-2+\sqrt{10}/2$ and $-2-\sqrt{10}/2.$ One can show that 
$$
e^{At}=P\left( \begin{array}{cc}
e^{(-2+\sqrt{10}/2)t} & 0 \\
0 & e^{(-2-\sqrt{10}/2)t} \end{array} \right)P^{-1},
$$
where
$P=\left( \begin{array}{cc}
1 & (-4+\sqrt{10})/3 \\
(-4+\sqrt{10})/2 & 1 \end{array} \right).$ Thus, the inequality $\left\|e^{At}\right\| \le K e^{-\omega t}$ is valid for $t \geq 0$ with $K=\left\|P\right\| \left\|P^{-1}\right\| \approx 2.0685$ and $\omega=2-\sqrt{10}/2.$

One can verify numerically that the solutions of (\ref{example_delay_system}) eventually enter the compact region $$\mathscr{D}=\left\{(x_{1}, x_{2}) \in \mathbb R^2 :~ 0.14 \leq x_{1} \leq 0.26, \ -0.06 \leq x_{2} \leq 0.05 \right\}$$ as $t$ increases. Therefore, it is reasonable to consider the conditions $(C1)$ and $(C2)$ inside the region $\mathscr{D}.$

Conditions $(C1)-(C3)$ are valid for system (\ref{example_delay_system}) with $M_{f}=0.001352$ and $L_{f}=0.0104.$ According to Theorem \ref{main_thm1}, system (\ref{example_delay_system}) possesses a unique uniformly exponentially stable unpredictable solution.

\begin{figure}[ht] 
\centering
\includegraphics[width=14.0cm]{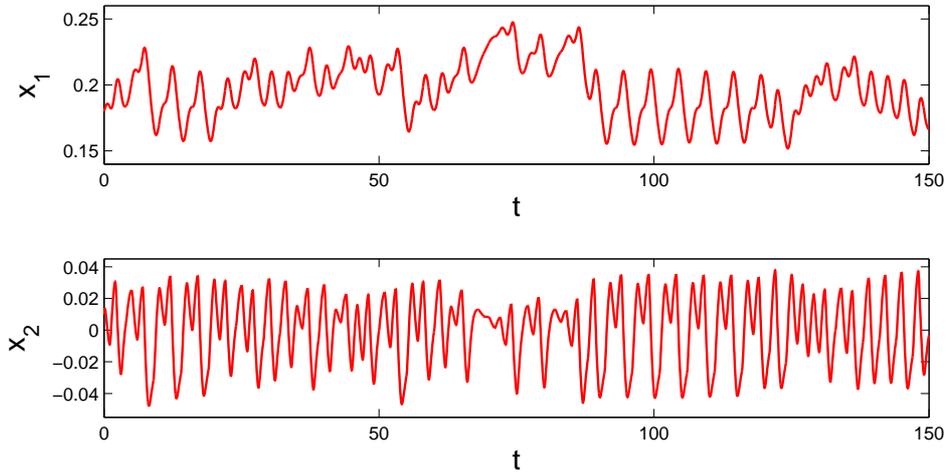}
\caption{The time series of $x_1$ and $x_2-$coordinates of system (\ref{example_delay_system2}). Chaotic behavior in both coordinates is observable in the figure.}
\label{fig2}
\end{figure}

\begin{figure}[ht] 
\centering
\includegraphics[width=10.0cm]{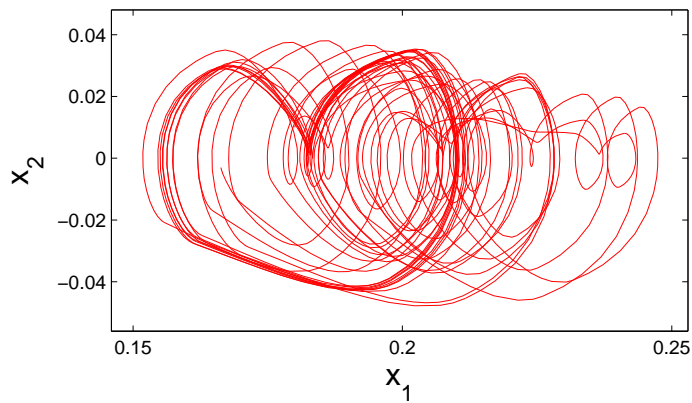}
\caption{The trajectory of system (\ref{example_delay_system2}). The figure manifests that the dynamics of (\ref{example_delay_system2}) is chaotic.}
\label{fig3}
\end{figure}

To demonstrate the chaotic dynamics of (\ref{example_delay_system}), let us consider the system
\begin{eqnarray} \label{example_delay_system2}
&& x'_1(t)=x_2(t), \nonumber \\ 
&& x'_2(t)=-1.5x_1(t)-4x_2(t)-0.02 x_1^2(t-0.1)+v(t),
\end{eqnarray}
where $v(t)$ is the solution of (\ref{system1_example}) represented in Figure \ref{fig1}. The $x_1$ and $x_2-$coordinates of the solution of (\ref{example_delay_system2}) corresponding to the initial conditions $x_1(t)=0.18,$ $x_2(t)=0.01,$ $t\in [-0.1,0]$ are shown in Figure \ref{fig2}. The figure supports the result of Theorem \ref{main_thm1} such that (\ref{example_delay_system}) possesses an unpredictable solution, and it reveals that the dynamics of (\ref{example_delay_system2}) is chaotic. Moreover, the trajectory of the same solution is depicted in Figure \ref{fig3}, and this simulation also confirms the presence of chaos in system (\ref{example_delay_system2}).

\section{Conclusion} 

Recently, we have introduced the concept of unpredictable points and motions for dynamical systems \cite{Akhmet16}. Next, an unpredictable function was defined as an unpredictable point of the Bebutov dynamics, and first theorems on the existence of unpredictable solutions were proved in \cite{Akhmet17,Akh18}. The metric of the Bebutov dynamics is not convenient for applications, since it is hard to verify. For this reason, in the present study, we apply the topology of convergence on compact sets to define unpredictable functions. The topology is metrizable and easy for applications with integral operators. Thus, one can accept that we lay a corner stone to the foundation of differential equations theory related to unpredictable solutions, and consequently, chaos.

% \section*{References}

%\bibliographystyle{model1-num-names}
%\bibliography{elsarticle-num.bst}

\end{document}